\documentclass[conference]{IEEEtran}
\usepackage{array}
\newcolumntype{C}[1]{>{\centering\arraybackslash}p{#1}}

\usepackage{algorithm}
\usepackage{algpseudocode}
\usepackage{cite}
\usepackage{amsmath,amssymb,amsfonts}
\usepackage{graphicx}
\usepackage{float}
\usepackage{stfloats} 
\usepackage{tikz}
\usepackage{url}

\usepackage{graphicx}
\usepackage[caption=false,font=footnotesize]{subfig} 
\usepackage{tikz}              
\usepackage[caption=false,font=footnotesize]{subfig} 
\usepackage{textcomp}
\usepackage{xcolor}
\def\BibTeX{{\rm B\kern-.05em{\sc i\kern-.025em b}\kern-.08em
    T\kern-.1667em\lower.7ex\hbox{E}\kern-.125emX}}
\makeatletter
\let\old@ps@headings\ps@headings
\let\old@ps@IEEEtitlepagestyle\ps@IEEEtitlepagestyle
\def\confheader#1{%
\def\ps@IEEEtitlepagestyle{
\old@ps@IEEEtitlepagestyle
\def\@oddhead{\strut\hfill#1\hfill\strut}
\def\@evenhead{\strut\hfill#1\hfill\strut}
}
\ps@headings
}
\makeatother
\confheader{
\small{Proceedings of the 13th RSI International Conference on Robotics and Mechatronics (ICRoM 2025), Dec. 16-18, 2025, Tehran, Iran} 
}
\usepackage[pscoord]{eso-pic}
\newcommand{\placetextbox}[3]{
\setbox0=\hbox{#3}
\AddToShipoutPictureFG*{ \put(\LenToUnit{#1\paperwidth},\LenToUnit{#2\paperheight}){\vtop{{\null}\makebox[0pt][c]{#3}}}
}
}
\placetextbox{.5}{0.055}{\textbf{\small{Proceedings of the 13th RSI International Conference on Robotics and Mechatronics (ICRoM 2025), Dec. 16-18, 2025, Tehran, Iran}}}

\begin{document}

\title{LSTM-Based Modeling and Reinforcement Learning Control of a Magnetically Actuated Catheter}
\author{
\IEEEauthorblockN{Arya Rashidinejad Meibodi}
\IEEEauthorblockA{
Advanced Service Robots (ASR) \\
Laboratory, Department of \\
Mechatronics Engineering, School \\
of Intelligent Systems Engineering,\\
College of Interdisciplinary Science \\
and Technology, University of Tehran\\
Tehran, Iran\\
aria.rashidi.nm@alumni.ut.ac.ir
}
\and
\IEEEauthorblockN{ Mahbod Gholamali Sinaki}
\IEEEauthorblockA{{Department of Mechanical Engineering} \\
\textit{K. N. Toosi University of Technology}\\
Tehran, Iran \\
sinaki@email.kntu.ac.ir}
\and
\IEEEauthorblockN{Khalil Alipour}
\IEEEauthorblockA{
Advanced Service Robots (ASR) \\
Laboratory, Department of \\
Mechatronics Engineering, School \\
of Intelligent Systems Engineering,\\
College of Interdisciplinary Science \\
and Technology, University of Tehran\\
Tehran, Iran\\
k.alipour@ut.ac.ir
}}\maketitle
\begin{abstract}
Autonomous magnetic catheter systems are emerging as a promising approach for the future of minimally invasive interventions. This study presents a novel approach that begins by modeling the nonlinear and hysteretic dynamics of a magnetically actuated
catheter system, consists of a magnetic catheter manipulated by servo-controlled magnetic fields generated by two external permanent magnets, and its complex behavior is captured using a Long Short-Term Memory (LSTM) neural network. This model validated against experimental setup's data with a root mean square error (RMSE) of 0.42 mm and 99.8\% coverage within 3 mm, establishing it as a reliable surrogate model. This LSTM enables the training of Reinforcement Learning (RL) agents for controlling the system and avoiding damage to the real setup, with the potential for subsequent fine-tuning on the physical system. We implemented Deep Q-Network (DQN) and actor-critic RL controllers, comparing these two agents first for regulation and subsequently for path following along linear and half-sinusoidal paths for the catheter tip. The actor-critic outperforms DQN, offering greater accuracy and faster performance with less error, along with smoother trajectories at a 10 Hz sampling rate, in both regulation and path following compared to the DQN controller. This performance, due to the continuous action space, suits dynamic navigation tasks like navigating curved vascular structures for practical applications.
\end{abstract}

\begin{IEEEkeywords}
Actor–critic, Deep Q-network (DQN), Deep reinforcement learning, Hysteresis modeling, Long short-term memory (LSTM), Magnetic catheter, Model-free control
\end{IEEEkeywords}

\section{Introduction}

Catheters, flexible tubular devices used in minimally invasive surgeries like cardiac ablation and neurovascular interventions, require precise navigation \cite{wang2024continuum, dupont2022continuum}. Traditionally, catheter navigation has relied on manual control by skilled clinicians or automated methods using magnetic actuation\cite{hwang2020review}, pneumatic systems\cite{ranzani2016soft}, tendon-driven mechanisms\cite{wang2025compact}, and hydraulic drives \cite{rich2018untethered}. Magnetic actuation, leveraging external permanent magnets to steer catheters, offers high precision but introduces complexities due to nonlinear and complex dynamics. Accurate modeling of these systems is challenging, particularly for magnetic continuum robots, where hysteresis(memory-dependent material behavior) complicates the relationship between control inputs and tip position \cite{bruder2020data},\cite{chen2024data}. Modeling catheter dynamics has historically involved physics-based approaches, such as finite element methods \cite{ferrentino2023finite}, Cosserat rod models \cite{tunay2011distributed}, pseudo-rigid-body models\cite{greigarn2017experimental}, and Euler-Bernoulli beam theory\cite{le2016accurate}, which struggle to capture catheter hysteresis due to inherent limitations, while methods like the Cosserat rod model also face significant computational complexity. Recent advancements in data-driven techniques, including neural networks, have improved environment reconstruction for soft continuum robot, with methods like neural network-based dynamic modeling\cite{liu2025data}. Recurrent neural networks, especially LSTMs, have proven especially effective in modeling temporal dependencies and hysteretic effects in robotic catheters, providing a robust solution for capturing nonlinear behaviors \cite{wu2021hysteresis,wang2024comparison, wang2024using}. However, controlling these complex systems poses a significant challenge due to their nonlinear dynamics, hysteresis, and susceptibility to external disturbances, such as varying magnetic fields or patient-specific anatomies. Conventional control methods, including Proportional-Integral-Derivative (PID) control, which operates without an explicit model but requires extensive tuning based on system response, and model predictive control (MPC), which relies on precise dynamic models, often struggle with robustness in unpredictable settings \cite{joseph2022metaheuristic}, \cite{morari1999model}. These methods face limitations such as poor adaptability to time-varying conditions, and difficulty handling unmodeled dynamics, necessitating advanced control strategies for magnetic catheter navigation\cite{bruder2020data}, \cite{wang2021survey}. RL offers a robust, model-free alternative, enabling agents to learn optimal control policies through trial and interaction with the environment \cite{10.5555/3312046, kargin2023reinforcement}. RL’s adaptability to dynamic, uncertain conditions makes it ideal for catheter navigation, where robustness in learning from limited data is critical \cite{kober2013reinforcement,scarponi2024zero}. This paper introduces a novel RL framework for magnetic catheter control, leveraging an LSTM-based plant to capture the exact dynamics, including hysteresis, of a continuum robot. We implement and compare two RL controllers: a DQN and an actor–critic algorithm based on Twin Delayed Deep Deterministic Policy Gradient (TD3), for point regulation and path-following tasks, achieving millimeter precision. Our novelty lies in the development of an LSTM model that accurately simulates the dynamics and hysteresis of the magnetically actuated catheter system, coupled with the application of both DQN and actor-critic deep RL techniques to achieve regulation and path following of the magnetic catheter's tip. Our results, demonstrate the actor-critic’s superiority in dynamic tasks, with potential for real-world deployment.
The remainder of this paper is organized as follows. 
Section~\ref{sec:Data Collection} describes the experimental setup and data acquisition protocol. 
Section~\ref{LSTM Modeling} details the dynamics and hysteresis modeling of the magnetically actuated catheter system using LSTM networks. 
Section~\ref{sec:RLCatheterControl} presents the design and implementation of the RL controllers. 
Section~\ref{Results and Discussion} reports and discusses simulation results. 
Finally, Section~\ref{sec:conclusion_future} presents the conclusions of this study and discusses potential avenues for future research.

\begin{figure*}[t]
\centering
\setlength{\tabcolsep}{5pt}
\begin{tabular}{cccc}
\begin{tikzpicture}[baseline=(img.north)]
  \node[inner sep=0] (img){\includegraphics[width=0.53\textwidth]{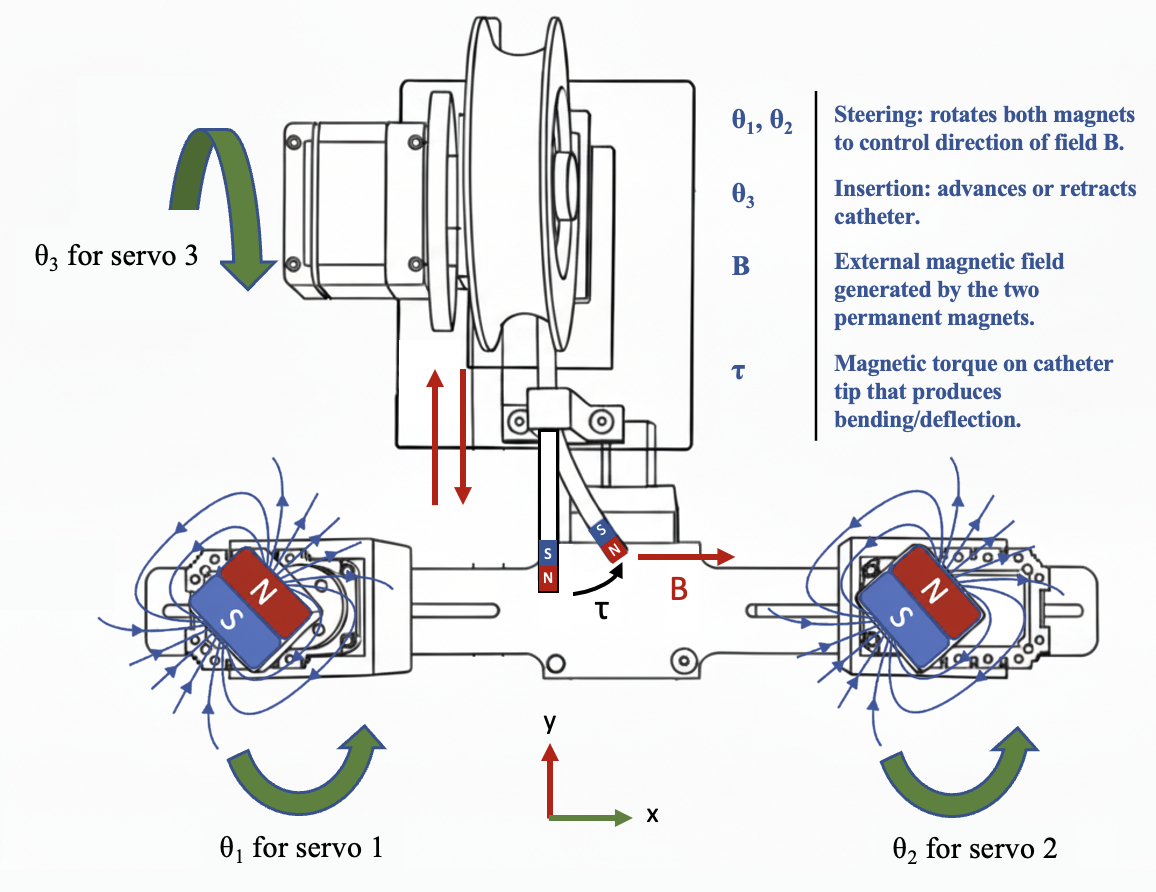}};
  \node[anchor=north, yshift=2pt] at (img.north west){\small\textbf{(a)}};
\end{tikzpicture}
&
\begin{tikzpicture}[baseline=(img.north)]
  \node[inner sep=0] (img){\includegraphics[width=0.40\textwidth]{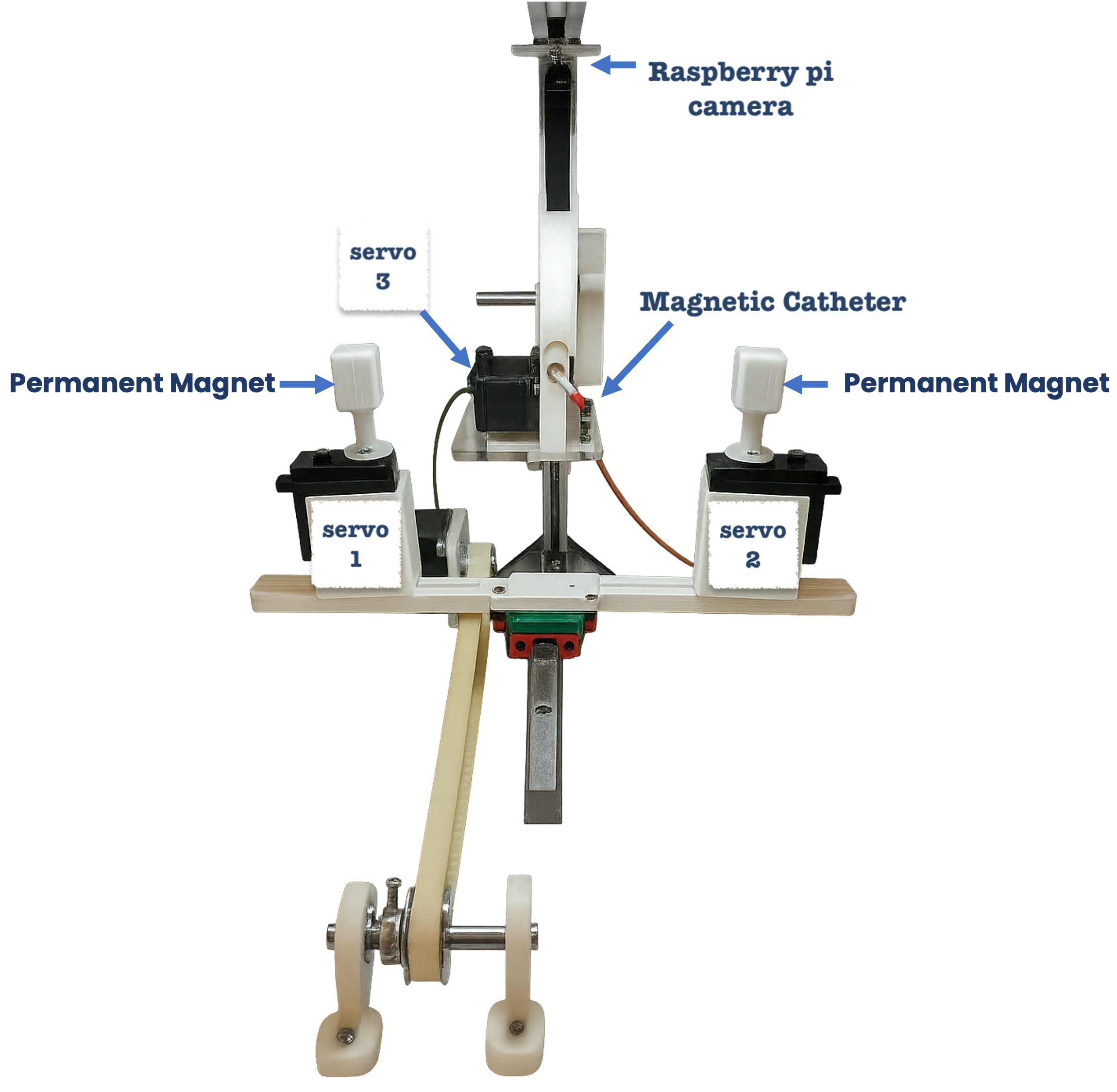}};
  \node[anchor=north, yshift=2pt] at (img.north west){\small\textbf{(b)}};
\end{tikzpicture}

\end{tabular}
\caption{(a) Top-view schematic of the magnetic catheter actuation system(\(\theta_{1}, \theta_{2}\) for steering, \(\theta_{3}\) for insertion)
and the resulting magnetic torque \(\tau\) on the catheter tip. (b) Experimental Configuration, comprising: Magnetic Catheter, Framework, Raspberry Pi along with Camera, External Permanent Magnets, and three Servo Motors.}
\label{fig:setup}
\end{figure*}

\section{Data Collection}
\label{sec:Data Collection}
\subsection{Experimental Setup}
\label{subsec:Experimental Setup}

A magnetic catheter with a length of $10 \,\text{cm}$ served as the central component of the experimental setup, characterized by physical attributes, including a Poisson's ratio of $0.3$, a Young's modulus of $200 \,\text{kPa}$, and a density of $1100 \,\text{kg/m}^3$. The distal tip of the catheter was equipped with four tiny magnets, designed to interact with external permanent magnets. Figure~\ref{fig:setup}(a) shows a schematic illustration of the magnetic catheter
actuation system. Two permanent magnets are mounted on Servo~1 and
Servo~2, which are mechanically coupled such that \(\theta_{2} = \theta_{1} +
180^\circ\). Rotating \(\theta_{1}\) simultaneously rotates both magnets, generating a external magnetic field
\(\mathbf{B}\) whose orientation can be arbitrarily chosen. The interaction between the tip magnet and the external field
\(\mathbf{B}\) produces a deflection torque \(\tau\) that bends the catheter tip in the
desired direction, thereby achieving desired steering. Servo~3 drives a mechanism that advances or retracts the entire catheter along its longitudinal
axis. The physical implementation is shown in Figure~\ref{fig:setup}(b). A Raspberry Pi camera with a resolution of \(1280\times720\) px, mounted above the setup and covering approximately \(100\,\text{mm}\) in the \(X\)-direction and \(120\,\text{mm}\) in the \(Y\)-direction in the image plane, was utilized to capture the catheter tip position.
The camera, integrated with OpenCV, tracked the \(X,Y\) coordinates of the catheter by detecting its red-colored tip, enabling real-time position data collection. Camera calibration with a planar checkerboard achieved a mean reprojection error of \(0.2\) px, corresponding to about \(0.02\text{--}0.03\,\text{mm}\) spatial uncertainty in the workspace, and repeated static-tip measurements showed position repeatability better than \(0.03\,\text{mm}\) in both axes. Images were captured at \(30\,\text{Hz}\) with short exposures (\(\leq 5\,\text{ms}\)), so motion blur and vision latency were negligible compared with the \(10\,\text{Hz}\) data sampling, while a wired control setup ensured accurate servo actuation and reliable catheter motion monitoring.

\subsection{Data Acquisition Protocol}

The data acquisition protocol was designed to capture the dynamic behavior of the magnetic catheter system under various operating conditions, providing comprehensive time-series data for LSTM model training. The methodology focused on collecting continuous motion data from the servo-actuated magnetic system while simultaneously recording the catheter's position response, with particular emphasis on characterizing the inherent hysteresis present in the catheter and magnetic actuation system. Data were collected at a constant sampling rate of \(10\,\text{Hz}\), which is safely above twice the highest system dynamics (approximately \(1\text{--}5\,\text{Hz}\)) and therefore satisfies the Nyquist–Shannon criterion, preventing aliasing. Sampling faster would mainly introduce redundant, highly correlated data without improving the surrogate model, and a \(5\text{--}20\,\text{Hz}\) range is consistent with sampling and control rates reported for related magnetic catheter and catheter-tracking systems\cite{tseng2019active, sun2025instant, hao2020contact, hao2025landing}. Each experimental run lasted between $10$--$30 \,\text{s}$, resulting in sequences of $100$--$300$ time steps per trial. Timestamps were recorded for each sample using the system clock, providing millisecond precision. 

The protocol focused on enhancing model robustness by incorporating diverse trajectories to address hysteresis in magnetic catheters, which stems from magnetic domain alignment, viscoelastic material properties, and frictional interactions with the medium. The experimental design included varied motion profiles to traverse the same servo angle space via ascending and descending paths, effectively capturing the system's history-dependent behaviors.

motion profiles for Servo 1,2 included:

\begin{itemize}
    \item \textbf{Linear Sweeps:} Servo1 was commanded to traverse its full range from $-175^{\circ}$ to $85^{\circ}$ at constant angular velocities ranging from $10^{\circ}/\text{s}$ to $30^{\circ}/\text{s}$, with Servo2 coupled with servo 1.
    These sweeps were conducted in both forward and reverse directions to capture the directional dependency of magnetic interactions, providing baseline data for analysis and revealing asymmetric responses characteristic of hysteresis.

    \item \textbf{Sinusoidal Trajectories:} Harmonic motion was implemented for Servo1's angle as follows:
    \begin{equation}
       \theta_{1}(t) = A \cdot \sin(2\pi f t + \phi)  
    \end{equation}
    
    where amplitude $A$ varied from $30^{\circ}$ to $90^{\circ}$, frequency $f$ ranged from $0.1 \,\text{Hz}$ to $1.0 \,\text{Hz}$, and phase $\phi$ was randomized between $0^{\circ}$ and $360^{\circ}$. The sinusoidal excitation is highly effective for characterizing hysteresis, as it creates closed loops by traversing the same angular positions in both ascending and descending phases, revealing variations in catheter positions based on motion direction and history.

    \item \textbf{Step Response Tests:} Discrete angular steps of $10^{\circ}$, $30^{\circ}$, and $60^{\circ}$ were applied to  servos, allowing characterization of transient responses and settling times. These tests were effective in observing immediate memory effects following sudden changes in magnetic field orientation, where the catheter's response lags due to magnetic domain realignment and mechanical damping.
\end{itemize}

The advancement mechanism (referred to as Servo3) was varied systematically across experiments. 
\begin{itemize}
\item \textbf{Discrete mode:} Servo3 positions were held constant at values of
0\textdegree,\allowbreak 8\textdegree,\allowbreak 16\textdegree,\allowbreak
24\textdegree,\allowbreak 32\textdegree,\allowbreak 40\textdegree,\allowbreak
48\textdegree,\allowbreak 56\textdegree,\allowbreak 64\textdegree,\allowbreak
72\textdegree,\allowbreak 80\textdegree, and 88\textdegree.
    \item \textbf{Continuous Mode:}
    Servo3 advanced at constant angular velocities of $10^{\circ}/\text{s}$ to $50^{\circ}/\text{s}$, were conducted in both
forward and reverse directions during servo motion, simulating realistic catheter navigation scenarios. 

\end{itemize}

 All data streams---servo commands, motor positions, and camera-detected coordinates---were synchronized using hardware timestamps and logged to CSV files with the format:
\[
[\text{timestamp}, \, \text{Servo1}, \, \text{Servo2}, \, \text{Servo3}, \, X, \, Y].
\]

This comprehensive dataset provided the foundation for subsequent LSTM modeling of the system's hysteresis and dynamic behavior.
\begin{figure}[t]
    \centering
    \includegraphics[width=1\linewidth]{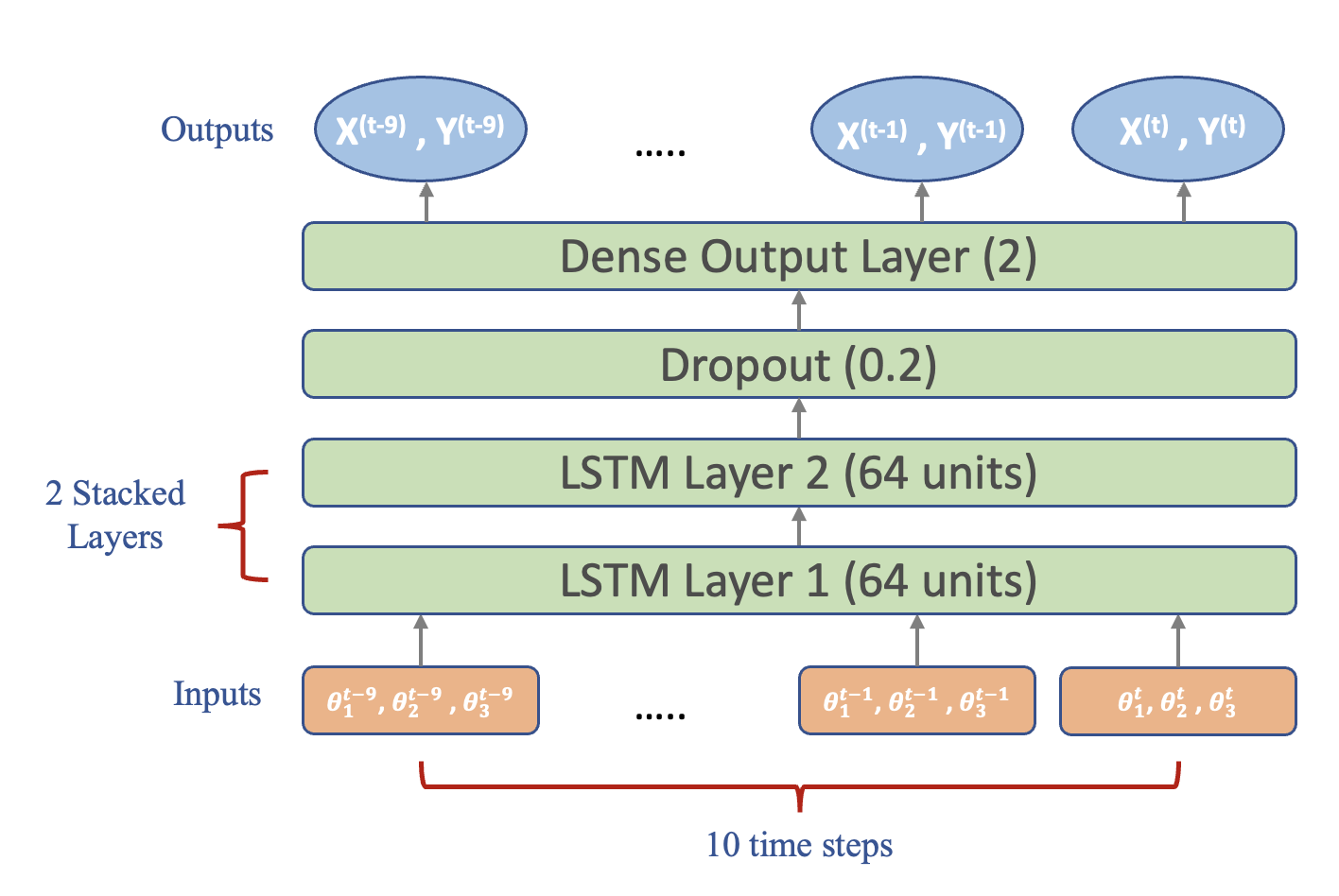}
    \caption{Schematic of the stacked LSTM network used for catheter tip position prediction. At each of the 10 time steps, the three normalized servo angles \((\theta_{1}, \theta_{2}, \theta_{3})\) are processed by two stacked LSTM layers (64 units each), a dropout layer (rate \(0.2\)), and a dense output layer that produces the corresponding normalized catheter tip coordinates \((X, Y)\).}
    \label{fig:lstm-architecture}
\end{figure}

\section{LSTM Modeling}
\label{LSTM Modeling}

\subsection{LSTM Model for Magnetic Catheter Prediction}

The LSTM model processes time-series data from the magnetic catheter system to predict catheter tip positions from servo actuation commands (see Fig.~\ref{fig:lstm-architecture}). The input consists of sequences of shape \([N, 10, 3]\), where \(N\) is the batch size, \(10\) represents one second of data at the \(10 \,\text{Hz}\) sampling rate, and \(3\) corresponds to the normalized servo angles \(\theta_{1}\), \(\theta_{2}\), and \(\theta_{3}\) (Servo1, Servo2, Servo3 respectively). The network architecture, illustrated in Fig.~\ref{fig:lstm-architecture}, comprises two stacked LSTM layers, each with \(64\) hidden units, followed by a dropout layer with rate \(0.2\) and a fully connected output layer. The LSTM processes the 10-step input sequences to produce hidden states \(h_t \in \mathbb{R}^{64}\) and cell states \(c_t \in \mathbb{R}^{64}\) at each time step. The final output layer transforms these hidden states to predict catheter positions, yielding output sequences of shape \([N, 10, 2]\) representing the normalized \(X\) and \(Y\) coordinates of the catheter tip. All inputs and outputs are normalized to the range \([0, 1]\) using Min–Max scaling based on the observed ranges from the experimental data.

The LSTM was trained on a comprehensive dataset consisting of approximately 160 structured experimental trials (linear sweeps, sinusoidal trajectories, step responses, and combined insertion motions) yielding approximately 30,000 raw samples. 
Key training hyperparameters are: batch size 128, Adam optimizer, an initial learning rate of 0.001 with a ReduceLROnPlateau scheduler (factor 0.5, patience 10, monitored on validation loss), and early stopping with a maximum of 300 epochs (patience 30 on validation loss). These hyperparameters were selected via grid search and cross-validation to balance convergence speed, generalization, and computational efficiency on our hardware. The internal LSTM dynamics follow the standard formulation of Hochreiter and Schmidhuber\cite{hochreiter1997long}.
Training was carried out on Google Colab (Tesla T4 GPU, 12 GB RAM) and took less than 12 minutes.

\subsection{LSTM Model Validation}
\label{sec:validation}

The LSTM model was validated on the real experimental plant to assess its accuracy in predicting catheter behavior. Validation experiments were conducted using a separate test set comprising 50 experimental runs, with a maximum of 240 time steps at 10 Hz, ensuring comprehensive coverage of the operating space. The model’s position predictions were compared against ground truth measurements obtained from a Raspberry Pi camera with OpenCV tracking. Three key performance metrics were assessed and are summarized in Table~\ref{tab:lstm_compact}. The coverage of predictions within specific error bands is detailed in Table~\ref{tab:lstm_coverage_style}, which highlights the model’s reliability across different thresholds. A sample comparison of real setup measurements and LSTM predicted values for the first 5 time steps at 10 Hz is provided in Table~\ref{tab:real_vs_pred}, illustrating the model’s performance over a short trajectory.

\begin{table}[!t]
  \caption{LSTM prediction error after inverse normalization.}
  \label{tab:lstm_compact}
  \centering
  \setlength{\tabcolsep}{3pt}%
  \renewcommand{\arraystretch}{0.95}%
  \scriptsize
  \begin{tabular}{@{}p{0.45\columnwidth}C{0.19\columnwidth}C{0.14\columnwidth}C{0.14\columnwidth}@{}}
    \hline
    \textbf{Metric} & \textbf{Overall [mm]} & \textbf{X [mm]} & \textbf{Y [mm]} \\
    \hline
    Root Mean Square Error (RMSE) & 0.42 & 0.38 & 0.45 \\
    Mean Absolute Error (MAE)     & 0.31 & 0.28 & 0.33 \\
    \hline
  \end{tabular}
\end{table}

\begin{table}[!t]
\caption{Coverage of predictions within absolute error bands.}
\label{tab:lstm_coverage_style}
\centering
\begin{tabular}{@{}p{0.35\columnwidth}C{0.35\columnwidth}p{0.2\columnwidth}@{}}
\hline
\textbf{Error band} & \textbf{Coverage} & \textbf{Notes} \\
\hline
$\le \pm 1\,\text{mm}$ & 92.3\% &  \\
$\le \pm 2\,\text{mm}$ & 98.7\% &  \\
$\le \pm 3\,\text{mm}$ & 99.8\% &  \\
Max $|$error$|$ [mm]   & 3.84 & worst case \\
\hline
\end{tabular}
\end{table}

\begin{table}[!t]
\caption{Real vs.\ LSTM predictions (first 5 steps at 10\,Hz).}
\label{tab:real_vs_pred}
\centering
\setlength{\tabcolsep}{3pt}%
\renewcommand{\arraystretch}{0.95}%
\scriptsize
\begin{tabular}{@{}
    C{0.13\columnwidth}  
    C{0.19\columnwidth}  
    C{0.19\columnwidth}  
    C{0.19\columnwidth}  
    C{0.19\columnwidth}@{}} 
\hline
\textbf{Time [s]} & \textbf{Real $X$ [mm]} & \textbf{Real $Y$ [mm]}
                  & \textbf{Pred.\ $X$ [mm]} & \textbf{Pred.\ $Y$ [mm]} \\
\hline
0.0 & 17.33 & 13.04 & 17.45 & 13.10 \\
0.1 & 17.52 & 12.80 & 17.60 & 12.85 \\
0.2 & 17.71 & 12.64 & 17.76 & 12.71 \\
0.3 & 17.90 & 12.44 & 17.85 & 12.50 \\
0.4 & 18.08 & 12.25 & 18.19 & 12.32 \\
\hline
\end{tabular}
\normalsize
\end{table}

The validation results demonstrated high accuracy of the LSTM predictions relative to the actual catheter behavior. To specifically validate the model's hysteresis modeling capability, we conducted targeted experiments comparing LSTM predictions for identical servo configurations reached through different motion histories. In these tests, the servos were commanded to reach the same configuration:
\begin{equation}
(\theta_1, \theta_2, \theta_3) = (-45^{\circ}, 135^{\circ}, 40^{\circ})
\end{equation}

via three distinct paths with $\theta_2 = \theta_1 + 180^{\circ}$ and $\theta_3$ constant at $40^{\circ}$:

\begin{enumerate}
    \item \textbf{Forward linear sweep:} $\theta_1: -175^{\circ} \rightarrow -45^{\circ}$ (increasing angle)
    \item \textbf{Reverse linear sweep:} $\theta_1: 85^{\circ} \rightarrow -45^{\circ}$ (decreasing angle)
    \item \textbf{Sinusoidal approach:} $\theta_1(t) = 60^{\circ} \sin(2\pi \cdot 0.5t)$, reaching $-45^{\circ}$ during the decreasing phase.
\end{enumerate}

The catheter tip positions for these three approaches were obtained from the experimental setup as $(X, Y) = (25.81, -17.12)$, $(24.31, -15.62)$, and $(26.51, -18.82)$ mm respectively, demonstrating clear hysteresis loops with position differences up to $2 \, \text{mm}$ in $X$ and $3 \, \text{mm}$ in $Y$. The LSTM model accurately predicted these differences, producing outputs of $(25.86, -17.15)$, $(24.21, -15.54)$, and $(26.43, -18.77)$ mm respectively, achieving prediction errors of less than $0.16 \, \text{mm}$ across all hysteresis branches.

This validation demonstrates the LSTM’s ability to capture path-dependent behavior and hysteresis through its cell state memory, effectively distinguishing the catheter tip position during increasing and decreasing servo angle trajectories despite identical final servo positions. These results establish the LSTM as a reliable forward model of the magnetically actuated catheter system, accurately capturing dynamics responses and hysteresis phenomena crucial for developing RL controllers.

\begin{figure*}[!t]
  \centering
  \includegraphics[width=\textwidth]{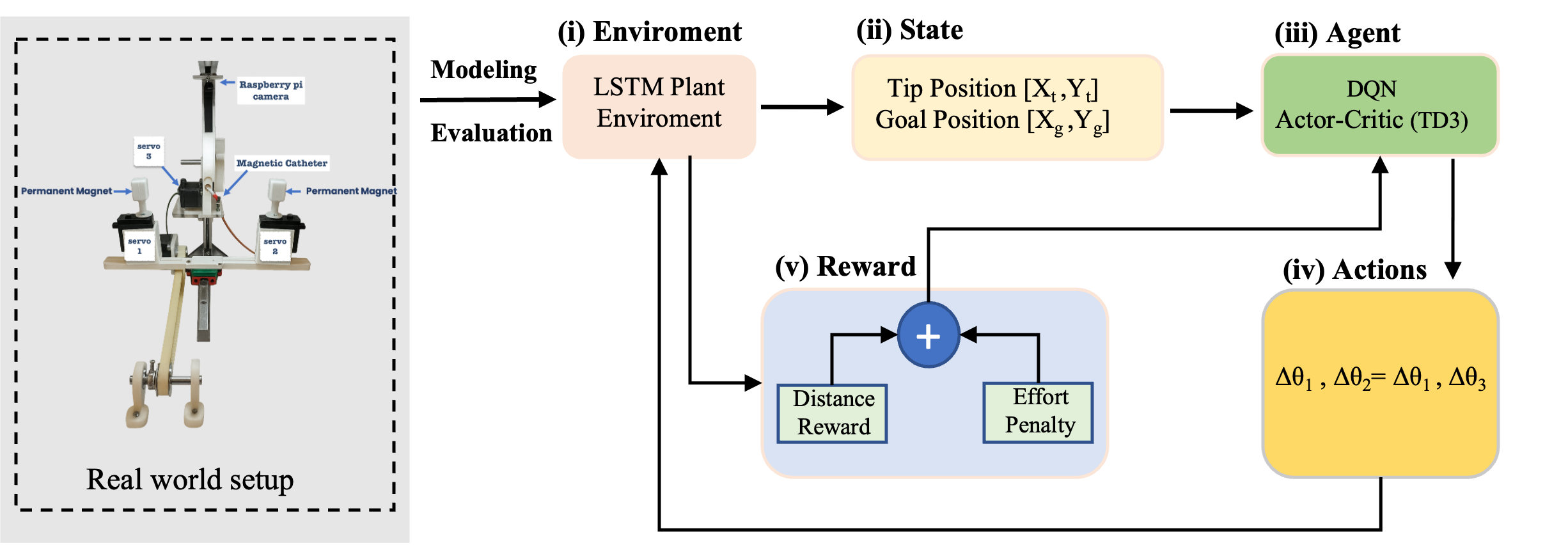}
  \caption{Schematic of the magnetic-catheter RL system. The real setup is modeled by an LSTM plant to form the RL environment (i).
At each step, the state (ii) contains the measured tip position \([X_t,Y_t]\) and the goal \([X_g,Y_g]\).
The agent (iii)—either a DQN or the actor–critic TD3 algorithm — selects actions (iv) as angular increments
\(\Delta\theta_{1},\, \Delta\theta_{3}\) (with \(\Delta\theta_{2}=\Delta\theta_{1}\) due to coupling).
The reward (v) is the sum of a goal-distance term and a control-effort penalty.
This loop is used for training and for closed-loop control of the magnetic catheter's tip position.
}
  \label{fig:wide_top}
\end{figure*}

\section{Reinforcement Learning for Catheter Control}
\label{sec:RLCatheterControl}

In this section, we introduce the RL based control system for the magnetic catheter (MC), employing deep RL techniques to guide the MC toward the target position, as depicted in Figure~\ref{fig:wide_top}.

\subsection{Simulation Environment}
\label{sec:Simulation Environment}

The simulation environment was developed in Python to replicate the magnetic catheter system's dynamics using the trained LSTM model as a surrogate plant. This model enables efficient training of reinforcement learning controllers without risking physical hardware damage or incurring experimental costs. The LSTM serves as the forward dynamics model, predicting catheter tip positions $[X, Y]$ from sequences of servo angles $[\theta_{1}, \theta_{2}, \theta_{3}]$, where $\theta_{2}$ is coupled to $\theta_{1}$ as $\theta_{2} = \theta_{1} + 180^{\circ}$.
The environment was implemented as a custom \texttt{CatheterEnv} class following the structure used in the provided implementation. At each simulation step (100~ms, matching the 10~Hz sampling rate), the controller provides an action that modifies the servo angles, and the LSTM processes the most recent 10-step input window to predict the corresponding 10-step position sequence. During the initial steps, when fewer than 10 data points are available, the missing inputs are zero-padded to maintain the fixed input length. To simulate real-world variability and prevent overfitting, zero-mean Gaussian noise ($\sigma = 0.3$~mm) was added to the predicted positions, promoting robust policy learning.

\subsection{Problem Definition and Components}
we employ two deep RL techniques, DQN and actor-critic, to address the catheter navigation control problem. The agent is trained to control the system and achieve desired behaviors by interacting with an LSTM-based simulation environment. The first step in formulating this control problem as an RL problem involves defining the state (observation) space, action space, reward function, and termination criteria, which are essential for structuring the learning task.

\subsubsection{State and Action Space Specification}

The state $\mathbf{s}_t = [X_t, Y_t, X_g, Y_g]^\top \in \mathbb{R}^4$ captures the current catheter tip position $[X_t, Y_t]$ and target position $[X_g, Y_g]$, both normalized to [0,1] via Min-Max scaling from experimental data ranges.

In every state, each servo angle ($\theta_1$, $\theta_2$, and $\theta_3$) can be updated by $\Delta\theta$. The action space varies by algorithm while respecting servo constraints:
\begin{equation}
\textbf{DQN (discrete actions):}\
\Delta\theta \in \{-5^\circ, -4^\circ, \dots, 4^\circ, 5^\circ\}^2
\end{equation}

\begin{equation}
\textbf{Actor-Critic(Continuous Actions): }  \Delta\theta \in [-5^\circ, +5^\circ]^2
\end{equation}

Actions are clipped to valid ranges: $\theta_1 \in [-175^\circ, 85^\circ]$, $\theta_2 \in [5^\circ, 265^\circ]$, $\theta_3 \in [0^\circ, 88^\circ]$. DQN's  discrete actions enable step-wise control, while actor-critic's continuous space facilitates smooth trajectories.

\subsubsection{Reward Function Design}

After applying an action, the environment returns a new state, and the reward is evaluated as:
\begin{equation}
r_t = - \left\| \mathbf{x}_t - \mathbf{x}_g \right\|_2 - \lambda \left( \lvert \Delta \theta_1 \rvert + \lvert \Delta \theta_2 \rvert +\lvert \Delta \theta_3 \rvert \right)
\end{equation}

where $\mathbf{x}_t$ is the current position of the catheter tip, $\mathbf{x}_g$ is the desired target catheter tip position, and $\lambda = 5 \times 10^{-3}$ is a regularization coefficient that balances position accuracy against control effort. The weighting factor \(\lambda = 5\times10^{-3}\) was selected empirically by trial-and-error: 
we varied \(\lambda\) over a small range and chose the value that produced accurate tracking with 
visibly smooth servo motions on validation episodes. This design prioritizes proximity to the target while penalizing excessive servo movements to ensure stability and efficiency.

\subsubsection{Episode Termination Criteria}

The final element to specify is the termination condition. An episode ends when the end effector's tip is sufficiently close to the goal or when the maximum episode length is exceeded. The term "sufficiently close" is checked by an adjustable threshold $\varepsilon$ based on the precision demanded:
\begin{equation}
\left\| \boldsymbol{x}_t - \boldsymbol{x}_{g} \right\|_2 < \varepsilon \quad \text{or} \quad t \geq T_{\max}
\end{equation}
where $\varepsilon$ is the distance threshold and $T_{\max}$ is the maximum time steps per episode (e.g., 150 steps or 15 seconds at 10 Hz).

\subsection{DQN Algorithm Implementation}

The Deep Q-Network (DQN) controller implements a value-based deep RL approach. A custom three-layer multilayer perceptron (MLP) with (128--128--$\text{ACTION\_DIM}$) units and ReLU activations is used to approximate the action-value function $Q(s,a)$. The discrete action space consists of 11 levels per servo (ranging from $-5^{\circ}$ to $+5^{\circ}$ in $1^{\circ}$ increments) for both  ($\Delta\theta_1$) and  ($\Delta\theta_3$), resulting in 121 possible action combinations ($\text{ACTION\_DIM} = 11 \times 11$). This formulation implicitly accounts for the $\theta_2$ coupling constraint. The implementation follows the provided code structure, employing experience replay (deque of 100{,}000 transitions), target network with soft updates ($\tau=0.005$), and $\varepsilon$-greedy exploration decaying from $\varepsilon = 1.0$ to $\varepsilon = 0.05$ over 6000 steps.

\subsection{Actor-Critic Algorithm Implementation} 
The actor-critic controller implements a policy-based deep RL approach using TD3 algorithm, tailored for continuous action spaces in catheter control. A three-layer multilayer perceptron (MLP) actor network (4-256-256-2 units) with ReLU activations and a \texttt{tanh} output layer approximates the policy function, producing continuous actions $\Delta \theta_1, \Delta \theta_3 \in [-5^\circ, +5^\circ]$. Twin three-layer MLP critic networks (6-256-256-1 units), also with ReLU activations, estimate the action-value function $Q(s, a)$ for state-action pairs, mitigating overestimation bias. This formulation accounts for the $\theta_2$ coupling constraint ($\theta_2 = \theta_1 + 180^\circ$) through coordinated action updates. The implementation follows the provided code structure, utilizing a replay buffer of 200,000 transitions, delayed policy updates (every 2 critic steps), and target network stabilization with soft updates ($\tau=0.005$). Exploration is enhanced with Gaussian noise ($\sigma=1^\circ$) added to actor outputs, decaying over episodes, while target policy smoothing applies Gaussian noise ($\sigma=0.25$) to improve stability. Training employs separate Adam optimizers (actor lr $=10^{-4}$, critics lr $=10^{-3}$), batch size 256, and discount factor $\gamma=0.99$, ensuring robust learning for the continuous servo control task.

\begin{figure*}[!t]
\centering

\subfloat{}{%
  \begin{tikzpicture}[baseline=(img.north)]
    \node[inner sep=0] (img) {\includegraphics[width=0.28\textwidth]{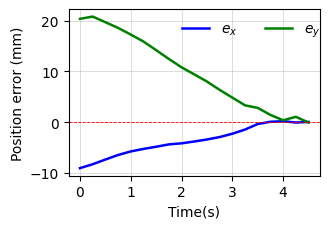}};
    \node[anchor=north east, xshift=-2pt, yshift=2pt] at (img.north west) {\small\textbf{(a)}};
  \end{tikzpicture}
}\hfill
\subfloat{}{%
  \begin{tikzpicture}[baseline=(img.north)]
    \node[inner sep=0] (img) {\includegraphics[width=0.28\textwidth]{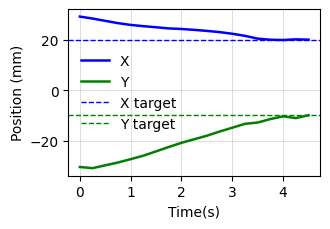}};
    \node[anchor=north east, xshift=-2pt, yshift=2pt] at (img.north west) {\small\textbf{(b)}};
  \end{tikzpicture}
}\hfill
\subfloat{}{%
  \begin{tikzpicture}[baseline=(img.north)]
    \node[inner sep=0] (img) {\includegraphics[width=0.28\textwidth]{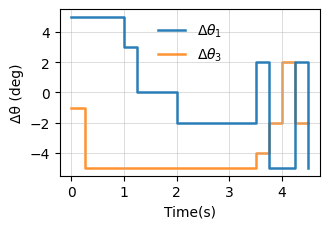}};
    \node[anchor=north east, xshift=-2pt, yshift=2pt] at (img.north west) {\small\textbf{(c)}};
  \end{tikzpicture}
}

\vspace{-1.2em}

\subfloat{}{%
  \begin{tikzpicture}[baseline=(img.north)]
    \node[inner sep=0] (img) {\includegraphics[width=0.28\textwidth]{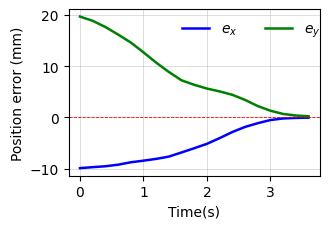}};
    \node[anchor=north east, xshift=-2pt, yshift=2pt] at (img.north west) {\small\textbf{(d)}};
  \end{tikzpicture}
}\hfill
\subfloat{}{%
  \begin{tikzpicture}[baseline=(img.north)]
    \node[inner sep=0] (img) {\includegraphics[width=0.28\textwidth]{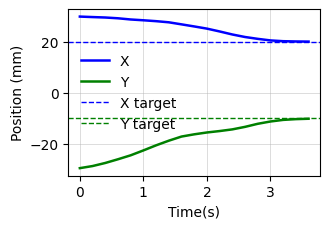}};
    \node[anchor=north east, xshift=-2pt, yshift=2pt] at (img.north west) {\small\textbf{(e)}};
  \end{tikzpicture}
}\hfill
\subfloat{}{%
  \begin{tikzpicture}[baseline=(img.north)]
    \node[inner sep=0] (img) {\includegraphics[width=0.28\textwidth]{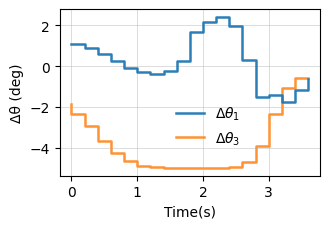}};
    \node[anchor=north east, xshift=-2pt, yshift=2pt] at (img.north west) {\small\textbf{(f)}};
  \end{tikzpicture}
}

\caption{Comparison of actor-critic and DQN controllers from initial state \([-10 \, \text{mm}, 20 \, \text{mm}]\) to target state \([20 \, \text{mm}, -10 \, \text{mm}]\). (a) errors of \(X\) and \(Y\) for DQN, (b) positions of \(X\) and \(Y\) for DQN, (c) \(\Delta \theta\) (actions) for DQN, (d) errors of \(X\) and \(Y\) for actor-critic, (e) positions of \(X\) and \(Y\) for actor-critic, and (f) \(\Delta \theta\) (actions) for actor-critic.}
\label{fig:comparison_six}
\end{figure*}
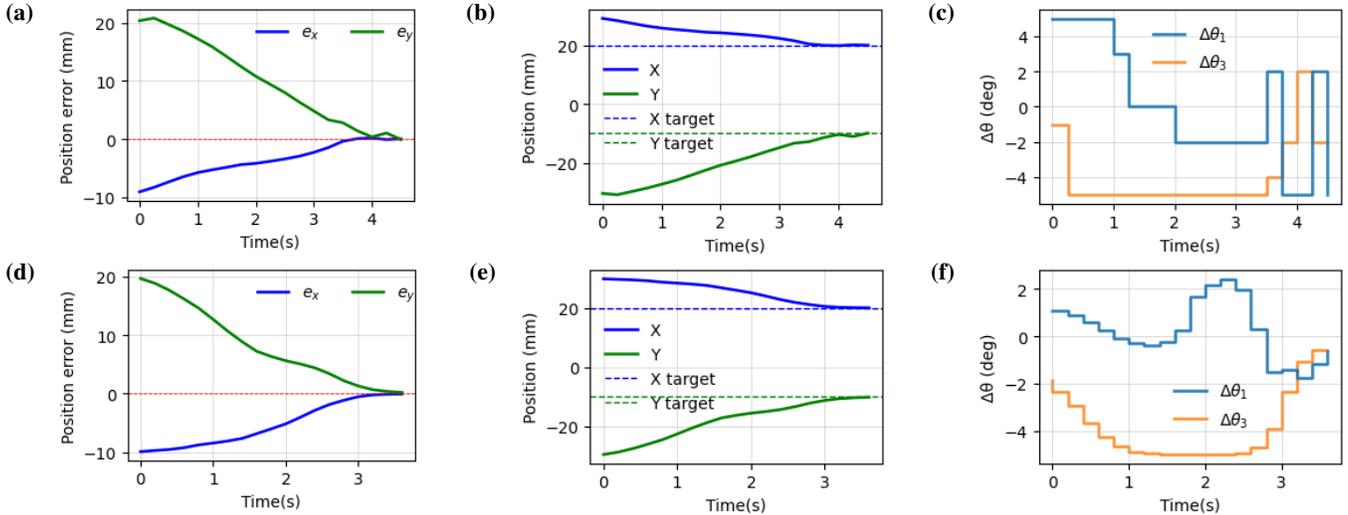

\section{Results and Discussion}
\label{Results and Discussion}
\subsection{Point Regulation Performance} The regulation performance of the DQN and actor--critic controllers was evaluated
for a fixed target position \([20\,\text{mm},\, -10\,\text{mm}]\), starting from
100 randomized initial states that were identical for both controllers. At a
sampling frequency of \(10\,\text{Hz}\), the DQN achieved a success rate of
\(98\%\), with an average of \(70.23\) steps to success and a mean final error of
\(0.170\,\text{mm}\). In contrast, the actor--critic achieved a \(100\%\) success
rate, with an average of \(61.420\) steps to success and a mean final error of
\(0.040\,\text{mm}\). These metrics, summarized in
Table~\ref{tab:batch_summary}, highlight the actor--critic's superior reliability
and precision. To better understand the performance of the two controllers, Figure ~\ref{fig:comparison_six} illustrates a trajectory from an initial state $[-10 \text{mm}, 20\text{mm}]$ to a target state $[20\text{mm}, -10\text{mm}]$, along with the $\Delta \theta$ (command) outputs of each controller. The results show that the actor-critic controller, with its continuous action space, produces smoother trajectories compared to the DQN controller. In this evaluation, a termination threshold of 0.02 mm was used; reducing this threshold causes the DQN controller to oscillate around the target, while the actor-critic controller maintains higher accuracy, handling thresholds as low as 0.005 mm. The chart is divided into three parts: first, the errors in $X$ and $Y$; second, the positions of $X$ and $Y$; and third, the $\Delta \theta$ (actions) for both controllers. The actor-critic's advantage stems from its continuous action adjustments, enabling finer control compared to the DQN. However, this comes at the cost of longer training time due to the complexity of policy optimization. With sufficient training, the actor-critic outperforms DQN in regulation tasks, offering greater accuracy and stability, which is critical for clinical applications where sub-millimeter precision is essential.

\begin{table}[!t]
\vspace{-4pt}
\caption{Evaluation over 100 randomized initial states. Goal $(20,-10)$~mm.}
\label{tab:batch_summary}
\centering
\scriptsize
\setlength{\tabcolsep}{4pt}%
\begin{tabular}{@{}p{0.43\columnwidth}C{0.25\columnwidth}C{0.25\columnwidth}@{}}
\hline
\textbf{Metric} & \textbf{DQN} & \textbf{Actor--Critic (TD3)} \\
\hline
Success rate (\%)                 & 98.0  & 100.0 \\
Avg.\ steps-to-success            & 70.23 & 61.42 \\
Avg.\ final $|e_x|$ (mm, succ)    & 0.045 & 0.014 \\
Avg.\ final $|e_y|$ (mm, succ)    & 0.158 & 0.037 \\
Avg.\ final $\lVert e \rVert$ (mm, succ) & 0.170 & 0.040 \\
\hline
\end{tabular}
\normalsize
\end{table}
\vspace{-4pt}

\subsection{Path Following Performance}

The controllers were further evaluated for path-following tasks using two reference paths: 

\begin{itemize}
    \item \textbf{Line path:} a straight segment from
    $A=(20,-10)$\,mm to $B=(30,-30)$\,mm:
    \begin{equation}
    X_{\text{line}}(t) = 20 + 10t,
    \end{equation}
    \begin{equation}
    Y_{\text{line}}(t) = -10 - 20t,
    \qquad t \in [0,1].
    \end{equation}

    \item \textbf{Half-sinus path:} a curved segment defined as
    \begin{equation}
    X_{\text{sin}}(t) = 20 + 8\sin(\pi t),
    \end{equation}
    \begin{equation}
    Y_{\text{sin}}(t) = -10 - 20t,
    \qquad t \in [0,1],
    \end{equation}
    discretized into $N=60$ waypoints.
\end{itemize}

\begin{figure*}[t]
\centering
\setlength{\tabcolsep}{5pt}
\begin{tabular}{cccc}
\begin{tikzpicture}[baseline=(img.north)]
  \node[inner sep=0] (img){\includegraphics[width=0.24\textwidth]{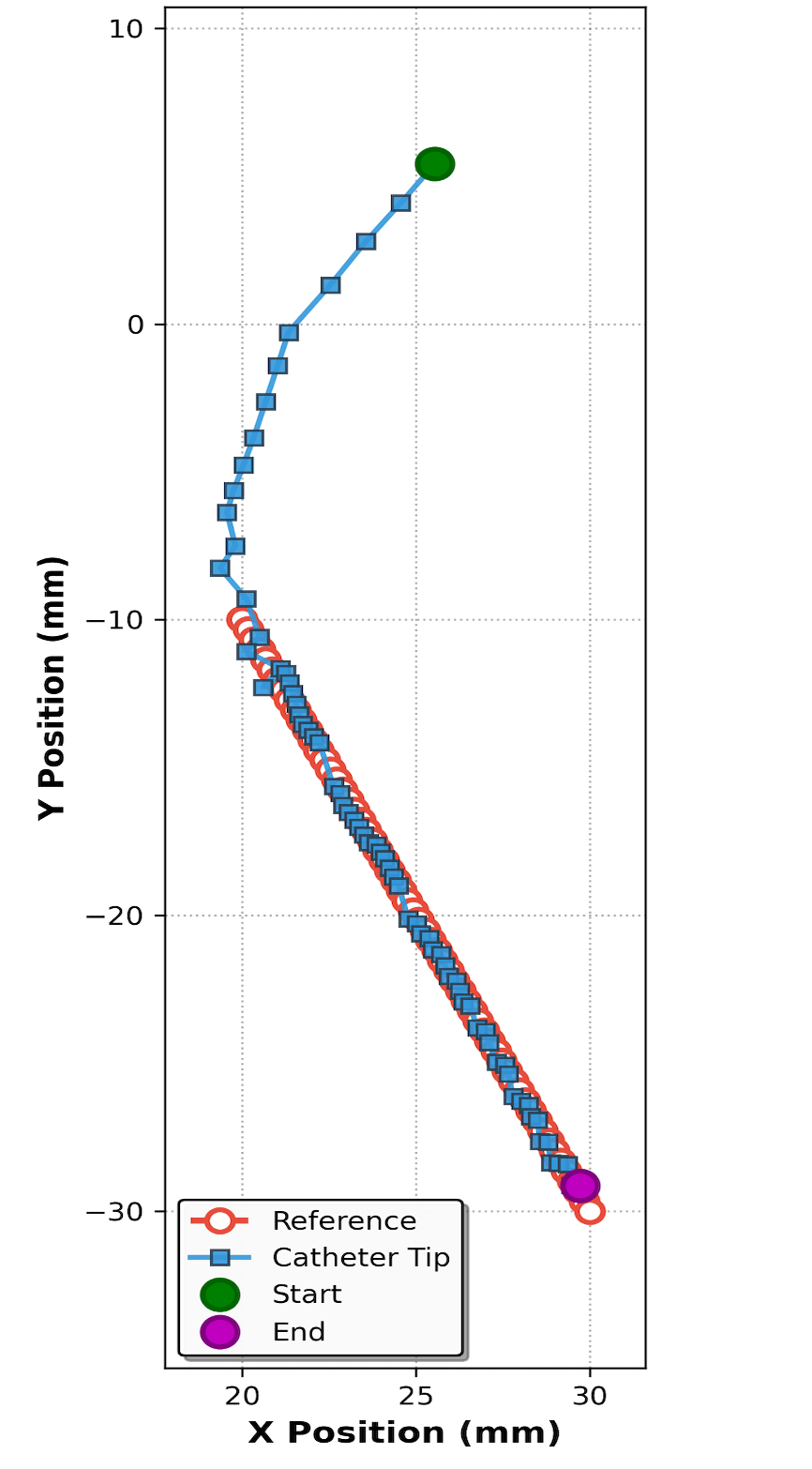}};
  \node[anchor=north, yshift=2pt] at (img.south){\small\textbf{(a)}};
\end{tikzpicture}
&
\begin{tikzpicture}[baseline=(img.north)]
  \node[inner sep=0] (img){\includegraphics[width=0.24\textwidth]{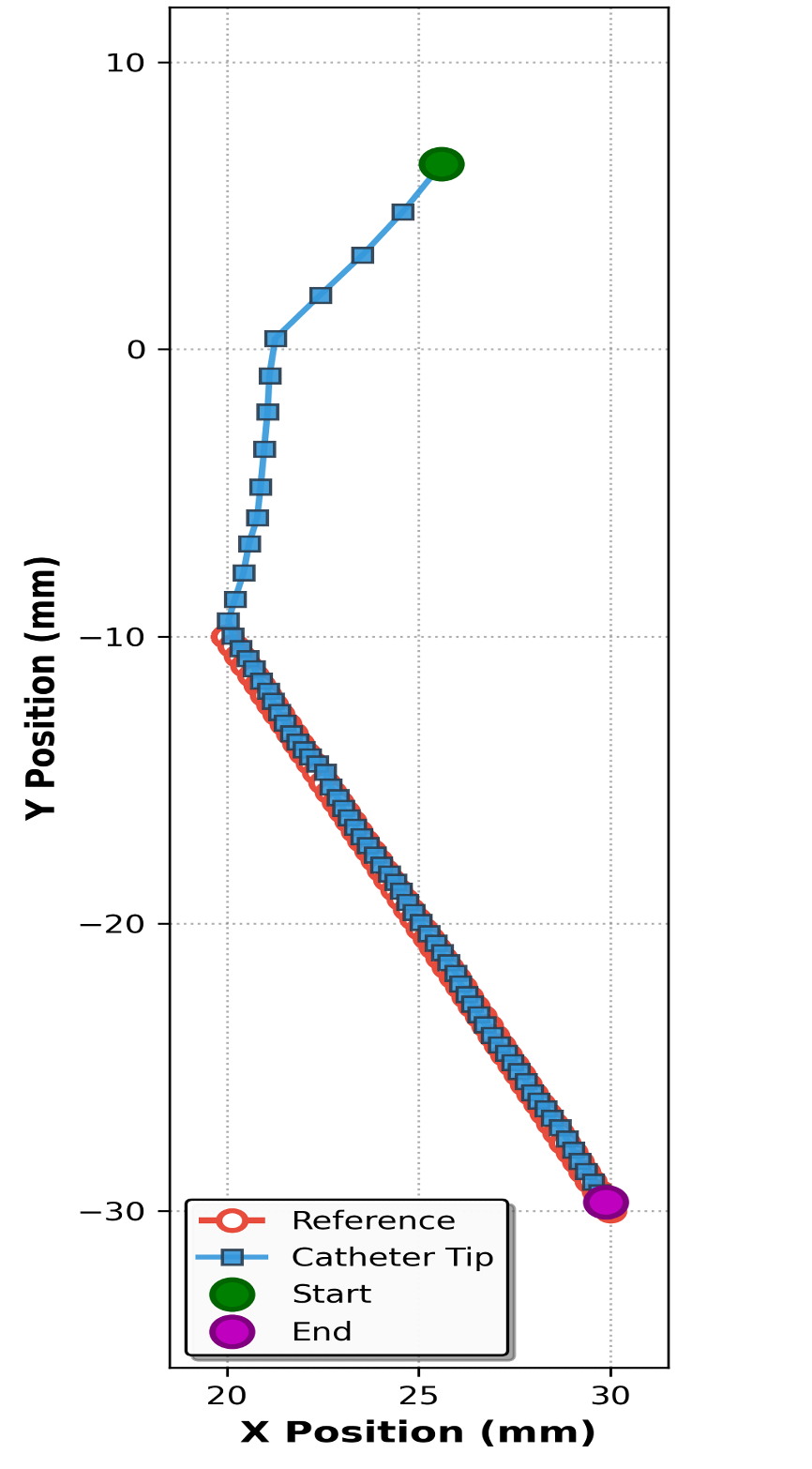}};
  \node[anchor=north, yshift=2pt] at (img.south){\small\textbf{(b)}};
\end{tikzpicture}
&
\begin{tikzpicture}[baseline=(img.north)]
  \node[inner sep=0] (img){\includegraphics[width=0.24\textwidth]{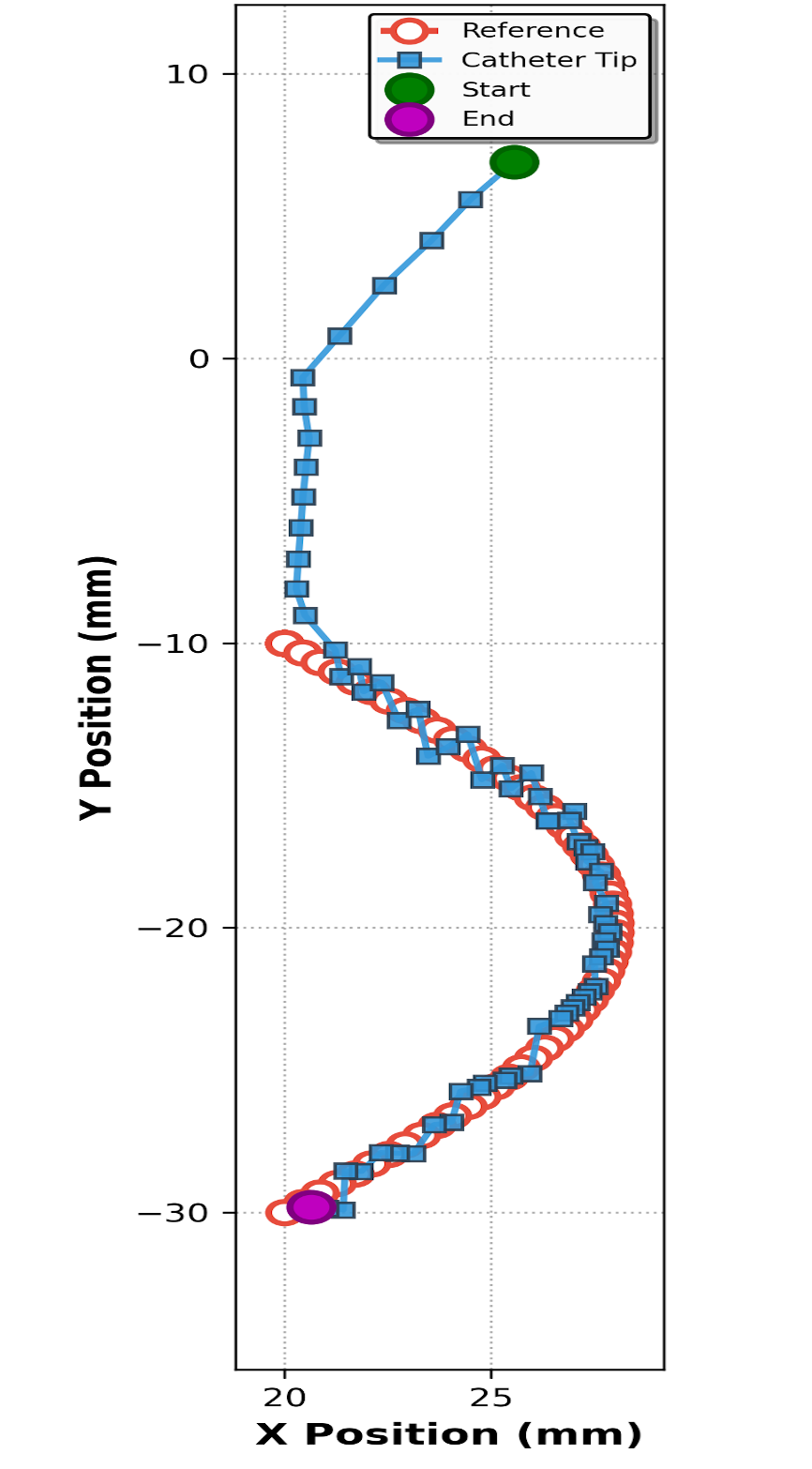}};
  \node[anchor=north, yshift=2pt] at (img.south){\small\textbf{(c)}};
\end{tikzpicture}
&
\begin{tikzpicture}[baseline=(img.north)]
  \node[inner sep=0] (img){\includegraphics[width=0.24\textwidth]{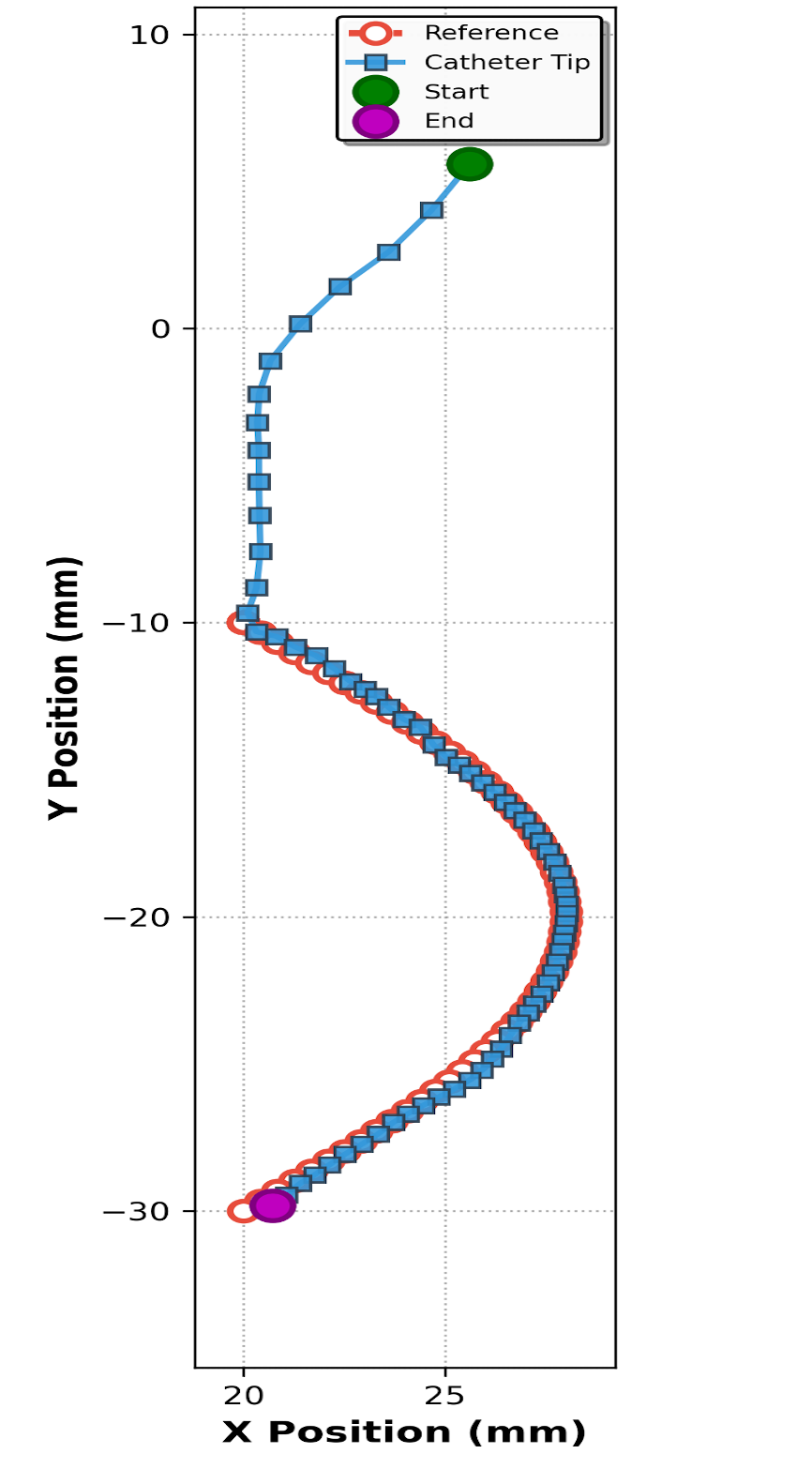}};
  \node[anchor=north, yshift=2pt] at (img.south){\small\textbf{(d)}};
\end{tikzpicture}
\end{tabular}

\caption{Path following performance on two reference paths: a straight line and a half-sinusoid. Red circles indicate the reference positions, while blue squares represent the catheter tip trajectories. (a) DQN for the straight-line path; (b) actor–critic (TD3) for the straight-line path; (c) DQN for the half-sinusoid path; (d) actor–critic (TD3) for the half-sinusoid path. Axes are in millimetres.}
\label{fig:comparison_four_two_by_two}
\end{figure*}
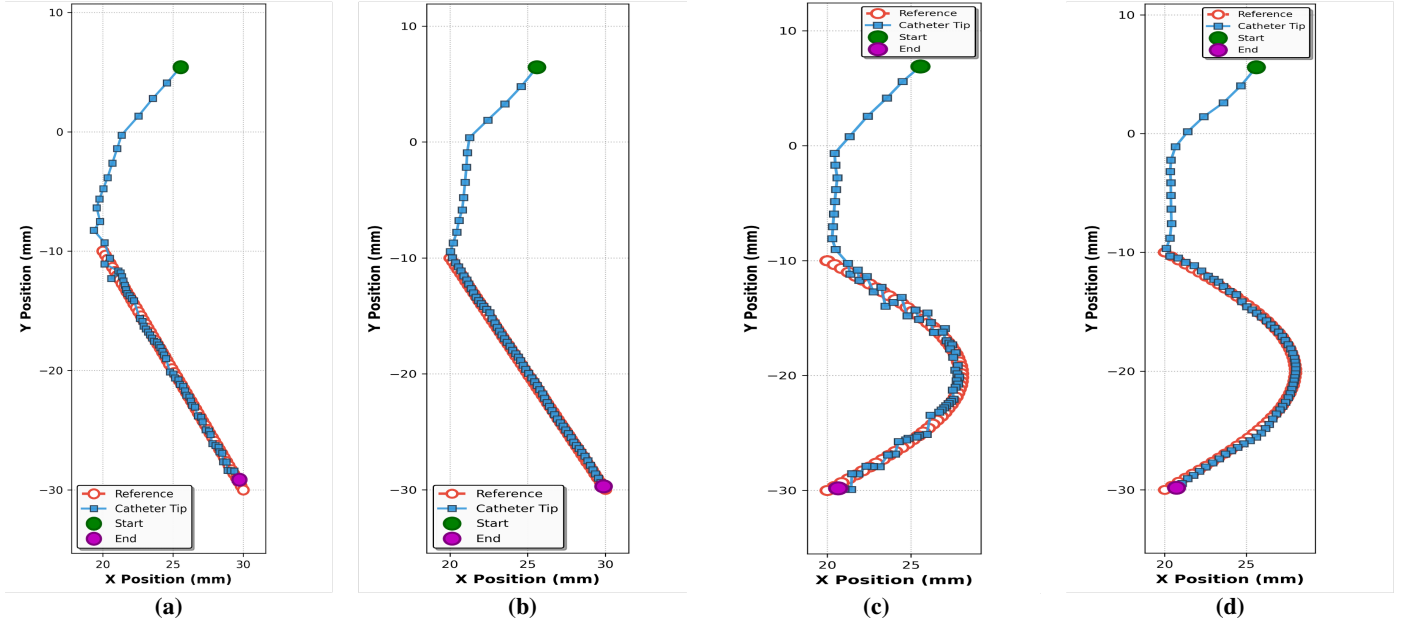

The controllers, trained to sequentially regulate the catheter tip to each waypoint, demonstrated robust tracking. 
Figure~\ref{fig:comparison_four_two_by_two} illustrates the performance of these controllers for both a linear and a half-sinusoidal path, with the plots showing the $X$ and $Y$ coordinates of the catheter tip versus the reference path.
Path Following accuracy was quantified using the mean path following error:
\begin{equation}
\overline{e} = \frac{1}{N}\sum_{i=1}^{N} \sqrt{(x_i - x_i^{\ast})^{2} + (y_i - y_i^{\ast})^{2}},
\end{equation}
where $(x_i, y_i)$ and $(x_i^{\ast}, y_i^{\ast})$ denote the catheter tip and reference positions, respectively, for each point along the path. For the \textbf{linear path}, the Actor--Critic controller achieved a mean error of $\overline{e}_{\mathrm{AC}} = 1.223~\mathrm{mm}$, while the DQN controller achieved $\overline{e}_{\mathrm{DQN}} =  1.898~\mathrm{mm}$. 
For the \textbf{half-sinusoidal path}, the mean errors were $\overline{e}_{\mathrm{AC}} = 1.187~\mathrm{mm}$ and $\overline{e}_{\mathrm{DQN}} = 2.547~\mathrm{mm}$. The DQN's performance was limited by its discrete action space, resulting in higher cumulative errors and less smooth trajectories, particularly along the curved sinusoidal path. The actor–critic, benefiting from its ability to adjust servo angles incrementally and continuously, reduced errors and maintained closer adherence to the reference paths. This smoothness and accuracy are attributed to the continuous action space making the actor-critic better suited for dynamic navigation tasks such as navigating curved vascular structures.

\section{Conclusion and Future Works}
\label{sec:conclusion_future}
We demonstrated that model-free deep RL controllers can effectively regulate the magnetic catheter tip position and achieve accurate path following without requiring a detailed model of its nonlinear continuum dynamics and magnetic field interactions. Additionally, we successfully modeled a highly nonlinear system with hysteresis using an LSTM, validated its accuracy through evaluation and comparison with experimental setup data, confirming its reliability as a highly accurate surrogate model. This enables the use of the LSTM for pre-training RL agents in simulation, avoiding potential damage to the real setup during training. The pre-trained agents can then be fine-tuned on the real-world setup, enhancing their adaptability. The actor-critic approach, outperforms DQN due to its continuous action space, achieving high precision with fewer discrete oscillations, and demonstrates capability in both point and path following. These advancements suggest potential applications in more practical contexts, such as clinical tests. Furthermore, we can use these agents and controllers on our experimental setup. Following fine-tuning and rigorous testing, we can deploy them in more practical applications, including more accurate vessel setups, clinical applications, and real-world scenarios.

\section*{Code availability}
The code used in this study has been deposited in a public GitHub repository and is available at:\\ 
\url{https://github.com/mahbos/MagneticCatheter-RL-LSTM}.

\section*{Acknowledgment}
We express our sincere gratitude to Dr. Hossein Nejat Pishkenari, Professor in the Department of Mechanical Engineering at Sharif University of Technology, for his invaluable guidance and contributions to this work. The experimental setup in his laboratory was used for modeling and simulation in this study.

\bibliographystyle{IEEEtran}
\bibliography{mybib}

@article{dupont2022continuum,
  title={Continuum robots for medical interventions},
  author={Dupont, Pierre E and Simaan, Nabil and Choset, Howie and Rucker, Caleb},
  journal={Proceedings of the IEEE},
  volume={110},
  number={7},
  pages={847--870},
  year={2022},
  publisher={IEEE}
}

@article{wang2025compact,
  title={Compact Design and Image-Space Pose Control of a Robot for Tendon-Driven Concentric Catheters in Mitral Repair Interventions},
  author={Wang, Weizhao and Xu, Zhouyang and Zeidan, Aya Mutaz and Saija, Carlo and Zheng, Yixuan and Arena, Matteo and Wang, Shuangyi and Housden, Richard James and Rhode, Kawal},
  journal={IEEE/ASME Transactions on Mechatronics},
  year={2025},
  publisher={IEEE}
}

@article{wang2021survey,
  title={A survey for machine learning-based control of continuum robots},
  author={Wang, Xiaomei and Li, Yingqi and Kwok, Ka-Wai},
  journal={Frontiers in Robotics and AI},
  volume={8},
  pages={730330},
  year={2021},
  publisher={Frontiers Media SA}
}

@article{ranzani2016soft,
  title={A soft modular manipulator for minimally invasive surgery: design and characterization of a single module},
  author={Ranzani, Tommaso and Cianchetti, Matteo and Gerboni, Giada and De Falco, Iris and Menciassi, Arianna},
  journal={IEEE Transactions on Robotics},
  volume={32},
  number={1},
  pages={187--200},
  year={2016},
  publisher={IEEE}
}

@article{hwang2020review,
  title={A review of magnetic actuation systems and magnetically actuated guidewire-and catheter-based microrobots for vascular interventions},
  author={Hwang, Junsun and Kim, Jin-young and Choi, Hongsoo},
  journal={Intelligent Service Robotics},
  volume={13},
  number={1},
  pages={1--14},
  year={2020},
  publisher={Springer}
}

@article{rich2018untethered,
  title={Untethered soft robotics},
  author={Rich, Steven I and Wood, Robert J and Majidi, Carmel},
  journal={Nature Electronics},
  volume={1},
  number={2},
  pages={102--112},
  year={2018},
  publisher={Nature Publishing Group UK London}
}

@article{bruder2020data,
  title={Data-driven control of soft robots using Koopman operator theory},
  author={Bruder, Daniel and Fu, Xun and Gillespie, R Brent and Remy, C David and Vasudevan, Ram},
  journal={IEEE transactions on robotics},
  volume={37},
  number={3},
  pages={948--961},
  year={2020},
  publisher={IEEE}
}

@article{chen2024data,
  title={Data-driven methods applied to soft robot modeling and control: A review},
  author={Chen, Zixi and Renda, Federico and Le Gall, Alexia and Mocellin, Lorenzo and Bernabei, Matteo and Dangel, Th{\'e}o and Ciuti, Gastone and Cianchetti, Matteo and Stefanini, Cesare},
  journal={IEEE Transactions on Automation Science and Engineering},
  volume={22},
  pages={2241--2256},
  year={2024},
  publisher={IEEE}
}

@article{ferrentino2023finite,
  title={Finite element analysis-based soft robotic modeling: Simulating a soft actuator in sofa},
  author={Ferrentino, Pasquale and Roels, Ellen and Brancart, Joost and Terryn, Seppe and Van Assche, Guy and Vanderborght, Bram},
  journal={IEEE robotics \& automation magazine},
  volume={31},
  number={3},
  pages={97--105},
  year={2023},
  publisher={IEEE}
}

@inproceedings{tunay2011distributed,
  title={Distributed parameter statics of magnetic catheters},
  author={Tunay, Ilker},
  booktitle={2011 Annual International Conference of the IEEE Engineering in Medicine and Biology Society},
  pages={8344--8347},
  year={2011},
  organization={IEEE}
}

@inproceedings{greigarn2017experimental,
  title={Experimental validation of the pseudo-rigid-body model of the MRI-actuated catheter},
  author={Greigarn, Tipakorn and Jackson, Russell and Liu, Taoming and {\c{C}}avu{\c{s}}o{\u{g}}lu, M Cenk},
  booktitle={2017 IEEE International Conference on Robotics and Automation (ICRA)},
  pages={3600--3605},
  year={2017},
  organization={IEEE}
}

@article{le2016accurate,
  title={Accurate modeling and positioning of a magnetically controlled catheter tip},
  author={Le, Vi NT and Nguyen, Nghia H and Alameh, Kamal and Weerasooriya, Rukshen and Pratten, Peter},
  journal={Medical physics},
  volume={43},
  number={2},
  pages={650--663},
  year={2016},
  publisher={Wiley Online Library}
}

@article{wu2021hysteresis,
  title={Hysteresis modeling of robotic catheters based on long short-term memory network for improved environment reconstruction},
  author={Wu, Di and Zhang, Yao and Ourak, Mouloud and Niu, Kenan and Dankelman, Jenny and Vander Poorten, Emmanuel},
  journal={IEEE Robotics and Automation Letters},
  volume={6},
  number={2},
  pages={2106--2113},
  year={2021},
  publisher={IEEE}
}

@article{joseph2022metaheuristic,
  title={Metaheuristic algorithms for PID controller parameters tuning: Review, approaches and open problems},
  author={Joseph, Stephen Bassi and Dada, Emmanuel Gbenga and Abidemi, Afeez and Oyewola, David Opeoluwa and Khammas, Ban Mohammed},
  journal={Heliyon},
  volume={8},
  number={5},
  year={2022},
  publisher={Elsevier}
}

@article{morari1999model,
  title={Model predictive control: past, present and future},
  author={Morari, Manfred and Lee, Jay H},
  journal={Computers \& chemical engineering},
  volume={23},
  number={4-5},
  pages={667--682},
  year={1999},
  publisher={Elsevier}
}

@article{kober2013reinforcement,
  title={Reinforcement learning in robotics: A survey},
  author={Kober, Jens and Bagnell, J Andrew and Peters, Jan},
  journal={The International Journal of Robotics Research},
  volume={32},
  number={11},
  pages={1238--1274},
  year={2013},
  publisher={SAGE Publications Sage UK: London, England}
}

@article{liu2025data,
  title={Data-Driven Methods for Sensing, Modeling and Control of Soft Continuum Robot: A Review},
  author={Liu, Jiaqi and Duo, Youning and Chen, Xingyu and Zuo, Zonghao and Liu, Yuchen and Wen, Li},
  journal={IEEE/ASME Transactions on Mechatronics},
  year={2025},
  publisher={IEEE}
}

@book{10.5555/3312046,
author = {Sutton, Richard S. and Barto, Andrew G.},
title = {Reinforcement Learning: An Introduction},
year = {2018},
isbn = {0262039249},
publisher = {A Bradford Book},
address = {Cambridge, MA, USA}
}

@article{wang2024continuum,
  title={Continuum robots and magnetic soft robots: From models to interdisciplinary challenges for medical applications},
  author={Wang, Honghong and Mao, Yi and Du, Jingli},
  journal={Micromachines},
  volume={15},
  number={3},
  pages={313},
  year={2024},
  publisher={MDPI}
}

@article{wang2024comparison,
  title={Comparison of Classical, Neural Network and Hybrid Models for Hysteretic Single-tendon Catheter Kinematics},
  author={Wang, Yuan and Dupont, Pierre E},
  journal={IEEE Robotics and Automation Letters},
  year={2024},
  publisher={IEEE}
}

@inproceedings{wang2024using,
  title={Using neural networks to model hysteretic kinematics in tendon-actuated continuum robots},
  author={Wang, Yuan and McCandless, Max and Donder, Abdulhamit and Pittiglio, Giovanni and Moradkhani, Behnam and Chitalia, Yash and Dupont, Pierre E},
  booktitle={2024 International Symposium on Medical Robotics (ISMR)},
  pages={1--7},
  year={2024},
  organization={IEEE}
}

@article{kargin2023reinforcement,
  title={A reinforcement learning approach for continuum robot control},
  author={Kargin, Turhan Can and Ko{\l}ota, Jakub},
  journal={Journal of Intelligent \& Robotic Systems},
  volume={109},
  number={4},
  pages={77},
  year={2023},
  publisher={Springer}
}

@article{scarponi2024zero,
  title={A zero-shot reinforcement learning strategy for autonomous guidewire navigation},
  author={Scarponi, Valentina and Duprez, Michel and Nageotte, Florent and Cotin, St{\'e}phane},
  journal={International Journal of Computer Assisted Radiology and Surgery},
  volume={19},
  number={6},
  pages={1185--1192},
  year={2024},
  publisher={Springer}
}

@article{hochreiter1997long,
  title={Long short-term memory},
  author={Hochreiter, Sepp and Schmidhuber, J{\"u}rgen},
  journal={Neural computation},
  volume={9},
  number={8},
  pages={1735--1780},
  year={1997},
  publisher={MIT press}
}

@article{tseng2019active,
  title={Active tracked intramyocardial catheter injections for regenerative therapy with real-time MR guidance: feasibility in the porcine heart},
  author={Tseng, Cheyenne CS and Wenker, Steven and Bakker, Maarten H and Kraaijeveld, Adriaan O and Dankers, Patricia YW and Seevinck, Peter R and Smink, Jouke and Kimmel, Scott and van Slochteren, Frebus J and Chamuleau, Steven AJ and others},
  journal={EuroIntervention},
  volume={15},
  number={4},
  pages={E336--E339},
  year={2019},
  publisher={EuroPCR}
}

@article{sun2025instant,
  title={Instant variable stiffness in cardiovascular catheters based on fiber jamming},
  author={Sun, Yi and Piskarev, Yegor and Hofstetter, Etienne H and Fischer, Cedric and Boehler, Quentin and St{\'a}rek, Zden{\v{e}}k and Nelson, Bradley J and Floreano, Dario},
  journal={Science Advances},
  volume={11},
  number={6},
  pages={eadn1207},
  year={2025},
  publisher={American Association for the Advancement of Science}
}

@inproceedings{hao2020contact,
  title={Contact stability analysis of magnetically-actuated robotic catheter under surface motion},
  author={Hao, Ran and Greigarn, Tipakorn and {\c{C}}avu{\c{s}}o{\u{g}}lu, M Cenk},
  booktitle={2020 IEEE International Conference on Robotics and Automation (ICRA)},
  pages={4455--4462},
  year={2020},
  organization={IEEE}
}

@article{hao2025landing,
  title={Landing control of a magnetically actuated robotic catheter on beating heart surface},
  author={Hao, Ran and Cavu{\c{s}}o{\u{g}}lu, M Cenk},
  journal={Scientific Reports},
  volume={15},
  number={1},
  pages={34581},
  year={2025},
  publisher={Nature Publishing Group UK London}
}

\end{document}